\documentclass[12pt]{iopart}
\usepackage{graphicx}
\usepackage{hyperref}
\usepackage{url}
\usepackage{color,xcolor}
\usepackage{epstopdf}

\begin{document}
\title[New iron-based multiferroics with improper ferroelectricity]{New iron-based multiferroics with improper ferroelectricity}
\author{Jin Peng,$^1$ Yang Zhang,$^1$ Ling-Fang Lin,$^1$ Lin Lin,$^2$ Meifeng Liu,$^{2,3}$ Jun-Ming Liu$^{2,4}$ and Shuai Dong$^{1,*}$}
\address{$^1$ School of Physics, Southeast University, Nanjing 211189, China}
\address{$^2$ Laboratory of Solid State Microstructures and Innovative Center of Advanced Microstructures, Nanjing University, Nanjing 210093, China}
\address{$^3$ Institute for Advanced Materials and School of Physics and Electronic Science, Hubei Normal University, Huangshi 435002, China}
\address{$^4$ Institute for Advanced Materials, South China Normal University, Guangzhou 510006, China}
\ead{sdong@seu.edu.cn}

\submitto{\JPD}
\vspace{2pc}
\begin{abstract}

In this contribution to the special issue on magnetoelectrics and their applications, we focus on some single phase multiferroics theoretically predicted and/or experimentally discovered by the authors in recent years. In these materials, iron is the common core element. However, these materials are conceptually different from the mostly-studied BiFeO$_3$, since their ferroelectricity is improper. Our reviewed materials are not simply repeating one magnetoelectric mechanism, but cover multiple branches of improper ferroelectricity, including the magnetism-driven ferroelectrics, geometric ferroelectric, as well as electronic ferroelectric driven by charge ordering. In this sense, these iron-based improper ferroelectrics can be an encyclopaedic playground to explore the comprehensive physics of multiferroics and magnetoelectricity. Furthermore, the unique characteristics of iron's $3d$ orbitals make some of their magnetoelectric properties quite prominent, comparing with the extensively-studied Mn-based improper multiferroics. In addition, these materials establish the crossover between multiferroics and other fields of functional materials, which enlarges the application scope of multiferroics.
\end{abstract}

\noindent{\it Keywords}: multiferroics, improper ferroelectricity, iron oxide, iron selenide, iron fluoride
\section{Introduction}

\subsection{What is improper ferroelectrics}

Ferroelectrics are important functional materials, playing an irreplaceable role in sensors, information storage, transducers, and other applications. Magnetic materials are even more extensively used, especially in the information storage area. Generally speaking, multiferroics denote a class of materials that combine these two different characteristics. More rigorous definition of a multiferroic material is simultaneous presentation of more than one primary ferroic order parameter in a single phase \cite{Schmid:Fe}. The coexisting and crossover between ferroelectricity and magnetism not only generate emergent physics, but also provide more functionalities for applications, e.g. electric control of magnetism. Therefore, the field of multiferroic materials and magnetoelectricity is quite attractive and great progress have been made since the beginning of new century \cite{Fiebig:Jpd,Cheong:Nm,Ramesh:Nm,Wang:Ap,Dong:Ap,Fiebig:Nrm}.

Based on the origin of ferroelectricity, ferroelectrics can be classified  into two families: proper ferroelectrics and improper ferroelectrics. The ferroelectrics that have been studied and applied to industries in the past century are basically proper ferroelectrics. Its ferroelectricity originate from ``ferroelectric active" ions (e.g. those with the $d^0$ configuration or with the $6s^2$ lone pair), as found in BiFeO$_3$, PbTiO$_3$, and Pb(Fe$_{1/2}$Nb$_{1/2}$)O$_3$. Although most of the proper ferroelectrics are not magnetic, there are exceptions, such as BiFeO$_3$\cite{Li:Npjqm}.

Improper ferroelectrics form an emerging area with the research upsurge of multiferroic materials. Most improper ferroelectrics are multiferroics, although not all. Improper ferroelectricity is not caused by ``ferroelectric active" ions, but the phase transitions of other order parameters, e.g. structural transition (the so-called geometric ferroelectricity), charge-ordering (the so-called electronic ferroelectricity), or magnetic ordering (the so-called magnetic ferroelectricity) \cite{Cheong:Nm}.

\begin{figure}
\centering
\includegraphics[width=0.96\textwidth]{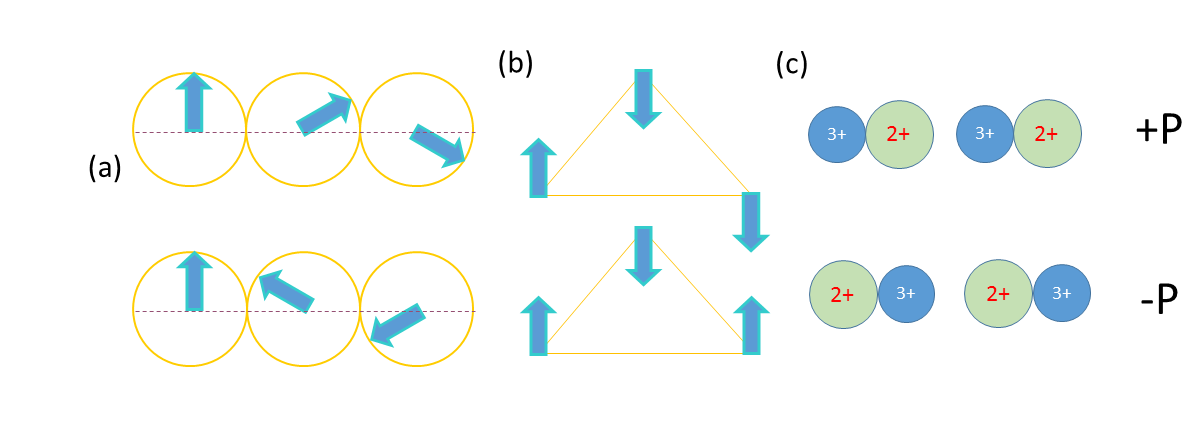}
\caption{Illustration of three types of improper multiferroics. (a) Magnetic ferroelectrics; (b) Geometric ferroelectrics; (c) Electronic ferroelectrics. The top and bottom panes denote the positive and negative ferroelectric state.}
\label{fig1}
\end{figure}

Magnetic ferroelectrics are achieved by some special magnetic orders, such as spiral magnetic ordering as shown in Fig.~\ref{fig1}(a). The displacement of electronic cloud and/or ions can be achieved via the spin-orbit coupling and/or spin-lattice coupling. These materials are also called the type-II multiferroics \cite{Khomskii:Phy}. The representative material is orthorhombic TbMnO$_3$ \cite{Kimura:Nat}.

Geometric ferroelectricity exists in some special lattices with geometric frustration, as shown in Fig.~\ref{fig1}(b). In these materials, ferroelectricity is generated by collaborative multiple nonpolar modes of lattice distortion.  Hybrid improper ferroelectrics, which have attracted great attentions recently, also belong to the category of geometric ferroelectrics \cite{Huang:Npjqm}. Some of these materials are nonmagnetic. However, since transition metal elements are mostly magnetic, most geometric ferroelectrics are multiferroics. The representative material is hexagonal YMnO$_3$ \cite{Aken:Nm,Pang:Npjqm}.

Electric dipoles and eventual ferroelectricity can be achieved via charge ordering, leading to so called electronic ferroelectrics \cite{Efremov:Nm}. Many transition metal ions have multiple valence states. In some special lattice environments, some elements can have ordered two valence states, forming the charge ordering, as illustrated in Fig.~\ref{fig1}(c). Some charge ordering can be switched between two ordered states, giving rise to ferroelectricity. The representative material is Pr$_{0.5}$Ca$_{0.5}$MnO$_3$ \cite{Efremov:Nm}.

These three types of improper ferroelectrics are mostly magnetic (although there are a few exceptions in geometric ferroelectrics, such as Ca$_3$Ti$_2$O$_7$), so they are mostly multiferroics, which generally have magnetoelectric coupling.

\subsection{Magnetoelectricity in improper multiferroics}
Magnetoelectricity is the correlation between magnetic moment and electric dipole moment. In particular, positive magnetoelectric effect means that the electric dipole moment can be controlled by magnetic field, while the inverse magnetoelectric effect denotes that the control of magnetic moment by an electric field. The magnetoelectric coupling effect has valuable applications, especially the inverse magnetoelectric effect. The control of magnetic moment by electric field can be energy conservative and efficient. It can overcome the technical bottleneck of current magnetic storage and ferroelectric storage. Magnetoelectric couplings are widely exist in single phase bulks, surfaces/interfaces, even in nonmagnetic and nonferroelectric topological insulators \cite{Dong:Ap}. However, the magnetoelectric coupling effects are very weak in most of these systems.

Multiferroics is an ideal platform for the pursuit of strong magnetoelectric coupling because of its intrinsic magnetism and electric dipoles. However, the general mutual exclusion between magnetic moment and electric dipole at the quantum level makes the pursuit of desirable multiferroic materials a challenging problem in condensed matter physics.

For those proper ferroelectrics with magnetism, such as BiFeO$_3$, the magnetic and ferroelectric properties originate from independent order parameters. Thus the Landau free energy of the phase transitions in these systems can be abstractly expressed as:
\begin{equation}
F(\textbf{P}, \textbf{L})=\alpha_{\rm Fe}\textbf{P}^{2}+ \beta_{\rm Fe}\textbf{P}^{4}+......+\alpha_{\rm AFM}\textbf{L}^2+ \beta_{\rm AFM}\textbf{L}^4+...,
\end{equation}
where $\textbf{P}$ is the ferroelectric order parameter; $\textbf{L}$ is the antiferromagnetic(AFM) order parameter; $\alpha$/$\beta$ are corresponding coefficients. Therefore, these systems can easily achieve good magnetic/ferroelectric properties. For example, BiFeO$_3$ has a large ferroelectric polarization ($\sim90$ $\mu$C/cm$^2$), high ferroelectric transition temperature ($\sim1100$ K), and high magnetic transition temperature($\sim660$ K) \cite{Wang:Sci}. However, due to the independency of magnetic and ferroelectric order parameters, these properties must be coupled by high-order effects, e.g. indirect and weak magnetoelectric coupling via lattice distortions. Recent theoretical work found that a small protion of BiFeO$_3$ polarization is improper\cite{Lee:Prl2015}. This kind of multiferroics can be considered as magnetoelectric composites in the atomic scale.

In contrast, it is hopeful to realize strong magnetoelectric coupling in improper ferroelectrics. The essence is to ``downgrade" the ferroelectric order parameter from the primary one to an dependent order parameter controlled by others.

In magnetic ferroelectrics represented by TbMnO$_3$, the magnetic order (mostly AFM order) is a primary order parameter, while ferroelectricity is a derivative of magnetic order. Therefore, the ferroelectric phase transition temperature is always equal to the magnetic ordering temperature. Thus ferroelectricity can be controlled by tuning magnetism.

In geometric ferroelectrics represented by hexagonal YMnO$_3$, the electronic polarization (distortion model $\Gamma^-_2$) is a derivative of the two collaborative non-polar lattice distortion modes ($K_1$: trimer of Mn triangular lattice; $K_3$: tilting of O$_6$ octahedral). Due to the strong order parameters of $K_1$ and $K_3$ modes,the ferroelectric transition temperature is very high (basically all above room temperature). Polarization is usually in the order of $10$ $\mu$C/cm$^2$. Meanwhile, the $K_3$ mode also controls the residual magnetic moment of canting antiferromagnetism. Therefore, the $K_3$ degree of freedom can regulate magnetism and electric properties simultaneously.Unlike proper ferroelectricity, the polarization of geometric ferroelectrics show hydrostatic pressure and thickness dependence in film sample\cite{Tan:JPCM2016,Xu:Prb2014,Sai:Prl2009}.

In electronic ferroelectrics represented by Pr$_{1/2}$Ca$_{1/2}$MnO$_3$, ferroelectricity is due to the coexistence of site-charge-ordering and bond-charge-ordering. It depends on the special charge ordered electronic configuration. Meanwhile, magnetism is also dependent on the electronic configuration. Therefore, the control of the electronic configuration can regulate magnetism and electric properties simultaneously.

In short, if the ferroelectric order parameter can be ``downgrade" to the derivative level of other order parameters, it may become easier to be controlled and more closely related to magnetism.

\subsection{Improper multiferroics: from Manganese-based to Iron-based}
As mentioned above, one can notice that many improper ferroelectrics are Mn-based oxides. The first magnetic ferroelectric material is TbMnO$_3$. The first geometric ferroelectrics is hexgonal YMnO$_3$, and the first candidate of electronic ferroelectrics is Pr$_{1/2}$Ca$_{1/2}$MnO$_3$. These Mn-based oxides cover all three known mechanisms of improper ferroelectricity. Study on these Mn-based oxides can establish the framework of improper ferroelectricity. Therefore, these materials have been extensively studied over the past fifteen years.

Why are the Mn-based oxides so ``magic"? First, Mn ions have multiple stable valence states in oxides: $+2$, $+3$, $+4$, as well as large magnetic moments correspondingly. For other $3d$ elements, Sc and Zn have neither multiple stable valence states nor magnetic moments; Ti, V, Co, and Ni do not have magnetic moment in all valence states.

Second, Mn has good chemical activity. Mn-based oxides display abundant crystal structures, including quasi-one dimensional, quasi-two dimensional, three-dimensional, square lattice and triangular lattice. It offers a fertile ground for improper ferroelectrics.

Third, the multiple stable valence states is the prerequisite for the formation of charge ordering, for example, Mn$^{3+}$/Mn$^{4+}$ can coexist in several kinds of systems.

Finally, the $3d$ orbital of Mn is the source of magnetic frustration. For example, in the $O_6$ octehedral, the $3d$ orbital splits into triplet $t_{\rm 2g}$ levels and doublet $e_{\rm g}$ levels. For the most common Mn$^{3+}$, $t_{\rm 2g}$ orbital is half occupied at the high spin state, and $e_{\rm g}$ is partially occupied. Therefore, $t_{\rm 2g}$ electrons prefer AFM interactions, and the $e_{\rm g}$ electrons prefer the ferromagnetic coupling. Variety of complex magnetic orders can be achieved via the competition between multiple exchanges \cite{Dagotto:Prp}.

This topical review will focus on iron-based improper ferroelectrics. Comparing with Mn-based improper ferroelectrics, the Fe-based ones own similar advantages.

\begin{figure}
\centering
\includegraphics[width=0.96\textwidth]{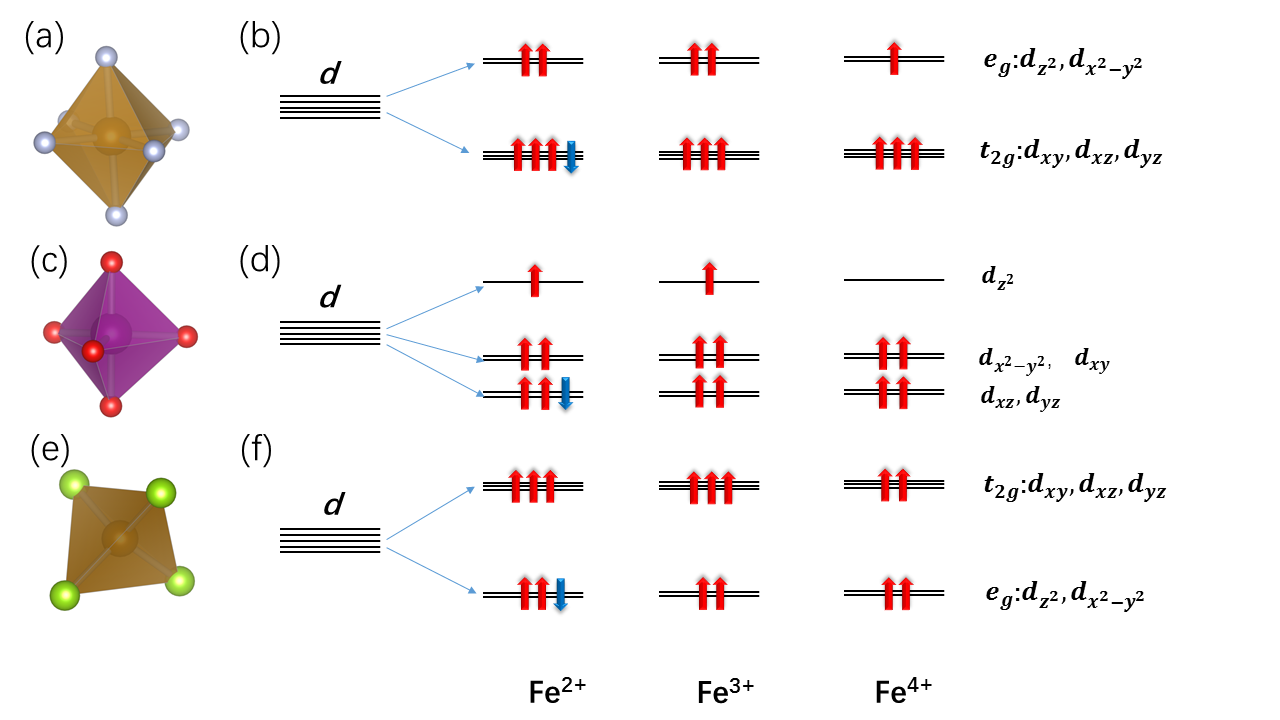}
\caption{Typical crystal environments of iron and its corresponding crystal field splitting of $3d$ orbital. (a) and (b): octahedral crystal field; (c) and (d): hexahedral crystal field; (e) and (f): tetrahedral crystal field. Spin up and spin down are represented by red and blue arrows respectively.}
 \label{fig2}
\end{figure}

First, iron also has multiple stable valence states in ionic crystals. The most common ones are $+2$ and $+3$ although $+4$/$+5$ also exist. Similar to Mn ions, Fe ions are generally at the high spin states, leading to large magnetic moments. Fe element also has good chemical activity and varies crystal structures. Typical crystal fields and corresponding electronic configurations are shown in Fig.~\ref{fig2}. It is not difficult to find out that these systems are also an ideal platform for the research of improper ferroelectrics. Iron-based improper ferroelectrics also cover geometric ferroelectrics, magnetic ferroelectrics, and electronic ferroelectrics.

The iron-based improper ferroelectrics also have several advantages compared with the Mn-based improper ferroelectrics. Iron's $3d$ orbital has stronger Coulomb repulsion $U$ than Mn. It has narrower $3d$ band and is much easier to form Mott insulating state. Enhanced correlation strength will also help to improve exchange interaction and enhance the magnetic transition temperature. For example, magnetic ordering temperature of rare-earth manganite $R$MnO$_3$ is generally below $200$ K \cite{Dagotto:Prp}, while the $R$FeO$_3$ become magnetic ordered above room temperature \cite{Hu:Csb}. From this point of view, the low common drawbacks of Mn-based improper ferroelectrics, such as high leakage and low magnetic ordering temperatures, may be solved in Fe-based compounds.

Also due to larger Coulomb repulsion $U$ and narrower $d$ band, charge ordered state is very common in iron-based compound. Further, different valence states (Fe$^{2+}$/Fe$^{3+}$ or Fe$^{3+}$/Fe$^{5+}$) are completely separated. In contrast, different valence states in manganese oxides (Mn$^{3+}$/Mn$^{4+}$) are not completely separated \cite{Coey:Nat}.

Last but not least, due to the discovery of iron-based superconductors, iron sulfides, iron selenides, iron pnictides and other non-oxides have been widely studied. Many new materials have been synthesized. These non-oxides greatly enrich the iron-based correlated electronic systems and provide a fertile ground for the study of improper ferroelectrics. By adjusting the anions, lattice structures, electronic structures and  magnetic properties can be well controlled.

\section{Iron-based multiferroics with improper ferroelectricity}
In the following, we will introduce several recently reported iron-based multiferroics with different magnetoelectric mechanisms. Although in all these compounds the element iron plays as the common leading role, these multiferroics cover almost all types of improper ferroelectricity, including geometric ferroelectricity, magnetic ferroelectricity, electronic ferroelectricity.

\subsection{Hexagonal LuFeO$_3$: geometric ferroelectricity}
Geometric ferroelectricity come from the structural instability in ionic crystals, similar to the conventional proper ferroelectricity. However, the driving forces are distinguishable between these two types of ferroelectricity. In  proper ferroelectrics like BaTiO$_3$, the polar phonon mode, i.e. the displacement of Ti$^{4+}$, is driven by forming a coordination bond between Ti$^{4+}$ and one of its neighbor O anions. The empty $3d$ orbital of Ti is crucial for this formation of coordination bond, implying the well-known $d^0$ rule for proper ferroelectricity \cite{Hill:Jpcb}. In contrast, in geometric ferroelectrics like hexagonal YMnO$_3$, the condensation of several non-polar phonon modes, i.e. trimerization of Mn triangles and tiltings of oxygen hexahedra, drives an uncompensated displacement of Y$^{3+}$ although neither Y$^{3+}$ nor Mn$^{3+}$ themselves is ferroelectric active \cite{Aken:Nm}. In other words, such improper ferroelectricity from geometric frustration, does not rely on the re-hybridization and covalency between ferroelectric-active cations and anions. Therefore, the geometric polar structure can be rather robust against carrier doping, and may persist even in the metallic materials \cite{Rondinelli:Am12}.

Hexagonal $R$MnO$_3$'s ($R$: rare earth or Y) usually have high ferroelectric transition temperatures ($T_{\rm C}\sim1000$ K), however, their AFM N\'eel temperatures ($T_{\rm N}$'s) only $\sim$ $100$ K \cite{filipetti:JMMM02}. The low magnetization temperature, as well as its large divergence with $T_{\rm C}$, prohibit strong magnetoelectric coupling and the application at ambient condition. As an alternative family, hexagonal $R$FeO$_3$'s also own geometric ferroelectricity, due to the same mechanism. It is expectable that stronger Fe$^{3+}$-Fe$^{3+}$ exchange interaction may enhance the magnetic ordering temperatures.

\begin{figure}
\centering
\includegraphics[width=0.96\textwidth]{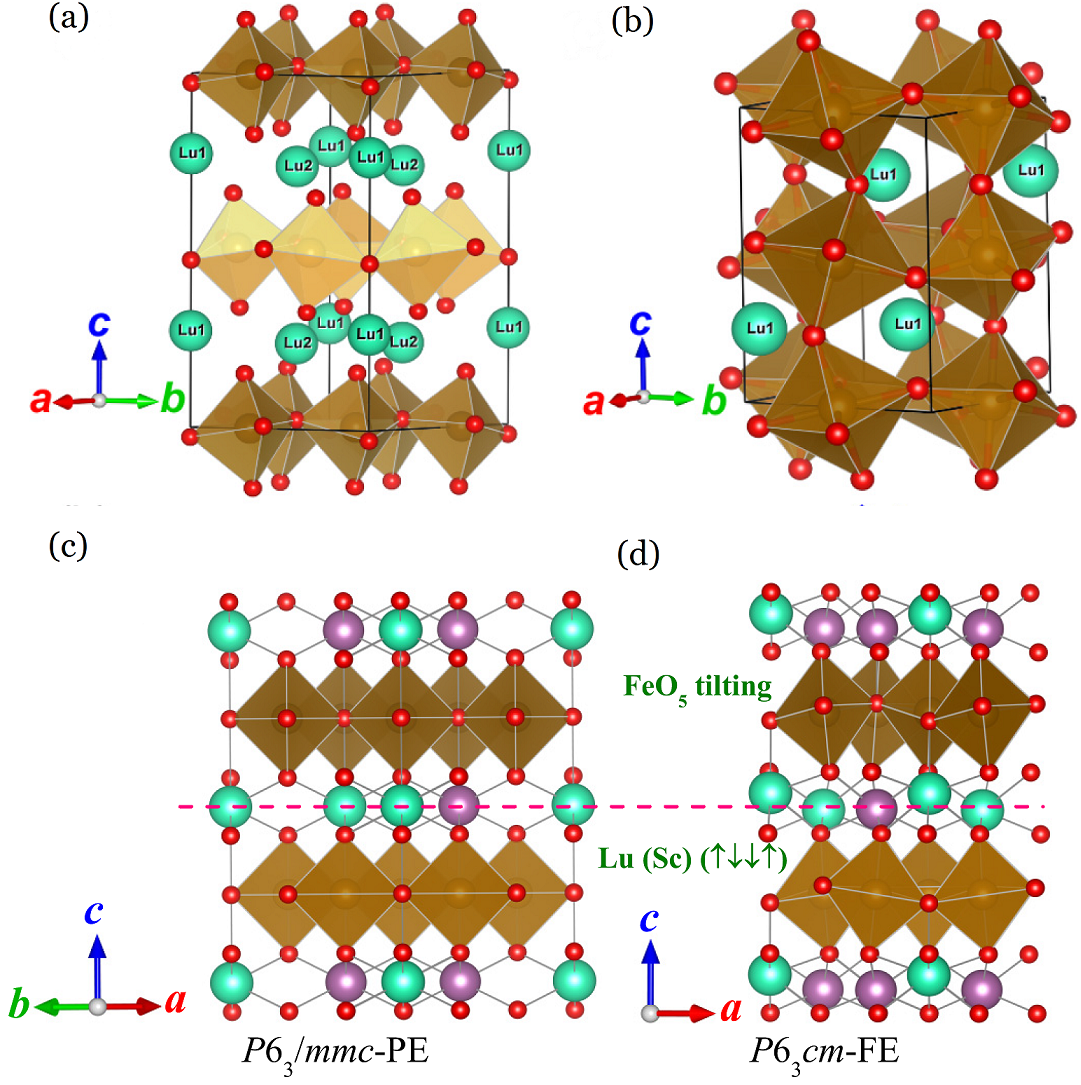}
\caption{(a) Crystal structure of hexagonal LuFeO$_3$. In $h$-LuFeO$_3$, each Fe$^{3+}$ are surrounded by five O, forming FeO$_{5}$ trigonal bipyramid. (b) Crystal structure of orthorhombic LuFeO$_3$. In $O$-LuFeO$_3$, each Fe$^{3+}$ are surrounded by O$_6$ octahedral. The crystal structure of (c) paraelectric and (d) ferroelectric states of $h$-Lu$_{0.5}$Sc$_{0.5}$FeO$_{3}$. Reprinted from Ref.~\cite{Lin:Prb16}, with the permission of American Physical Society.}
\label{LSFO1}
\end{figure}

$R$FeO$_3$ can crystallize in both the orthorhombic ($o$-$R$FeO$_3$) structure and hexagonal ($h$-$R$FeO$_3$) structure, while the orthorhombic one is the stablest at ambient condition. Taking the LuFeO$_3$ for example, its orthorhombic structure with the space group $Pbnm$ is non-polar (as shown in Fig.~\ref{LSFO1}(b)) and exhibits C-type antiferromagnetism below $620$ K \cite{zhu:APL12}. In contrast, a polar structure (space group $P6_{3}cm$, as shown in Fig.~\ref{LSFO1}(a)) has also been found in bulk (meta-stable) and thin films \cite{Xu:Mplb}. In this hexagonal phase, the paraelectric (space group $P6_3/mmc$) to ferroelectric transition at $\sim1050$ K can be achieved via the freezing of three phonon modes $\Gamma_2^-$, $K_1$, and $K_3$ as shown in Fig.~\ref{LSFO1}(c-d) \cite{Aken:Nm}. Distinct with the well-recognized ferroelectricity, the magnetism of $h$-LuFeO$_3$ was under debate. First, Wang \textit{et al.} reported an AFM order below $440$ K, followed by a spin reorientation resulting in a weak ferromagnetic order below $130$ K on $h$-LuFeO$_3$ thin film \cite{Wang:Prl}. However, a latter work by Disseler \textit{et al.}, could not confirm the high-temperature antiferromagnetism, while only the low-temperature transition at $\sim120-147$ K was reported \cite{Disseler:Prl}.

The metastability of bulk $h$-LuFeO$_3$ phase makes the comprehensive study of its magnetism quite challenging. Recently, several groups reported that a stable hexagonal structure can be achieved in scandium (Sc)-substituted LuFeO$_3$. At the half-substituted compounds (Lu$_{0.5}$Sc$_{0.5}$)FeO$_3$, pure hexagonal bulk phase can be synthesized \cite{masuno:IC13,disseler:PRB15}. Both the M$\ddot{o}$ssbauer spectrum and X-ray photoelectron spectroscopy (XPS) results suggest that the Fe ion exists as Fe$^{3+}$ in this composition. First-principles calculations show that the hexagonal structure can be stabilized by partial Sc substitution, while the multiferroic properties, including the noncollinear magnetic order and geometric ferroelectricity, remain robustly unaffected \cite{Lin:Prb16}. Therefore, Lu$_{1-x}$Sc$_x$FeO$_3$ can act as an alternative material to check the multiferroicity of LuFeO$_3$ and related materials in the bulk form.

\begin{figure}
\centering
\includegraphics[width=0.96\textwidth]{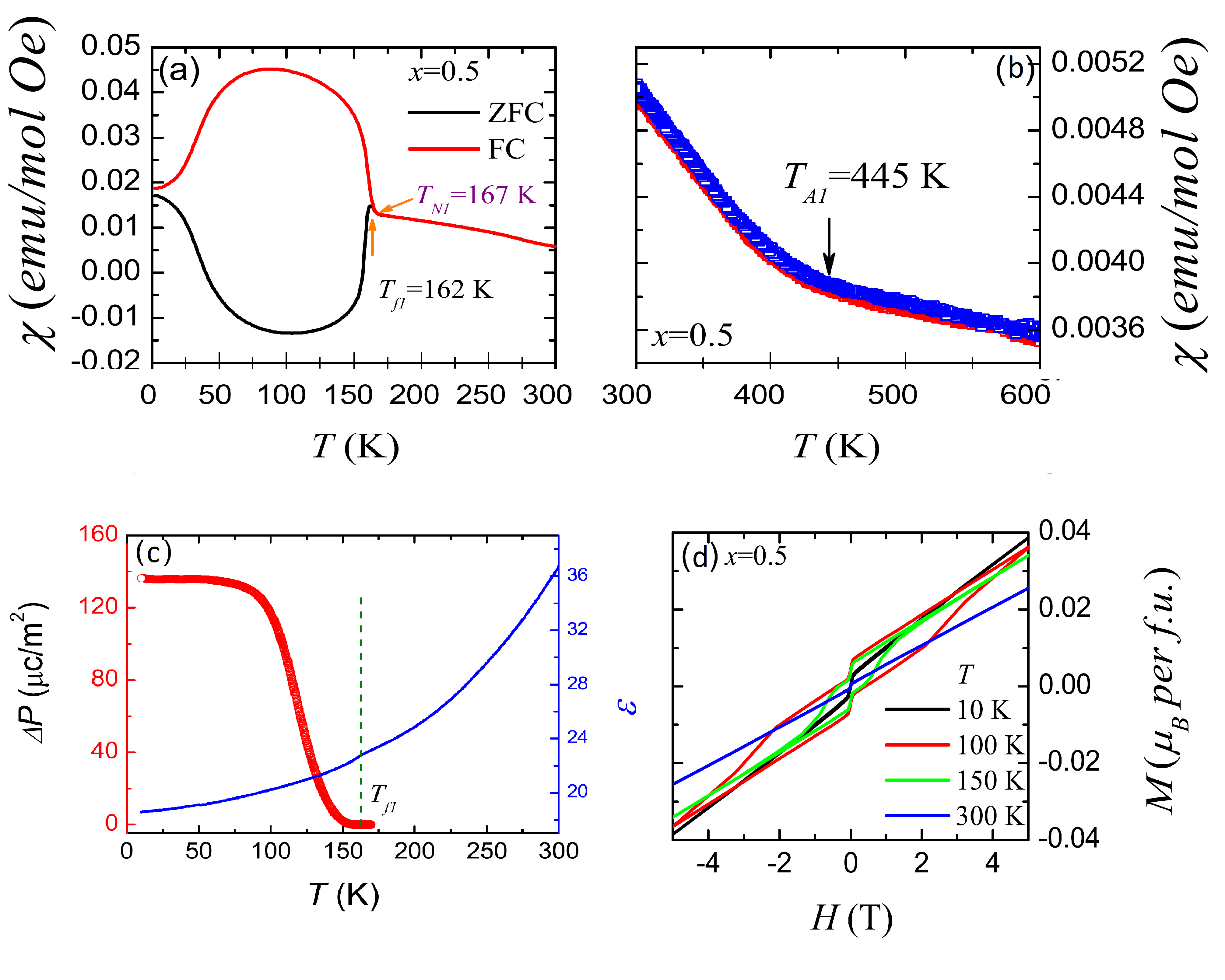}
\caption{(a) Temperature dependence of magnetic susceptibility $\chi$ of sample Lu$_{1-x}$Sc$_x$FeO$_3$ $x=0.5$ under zero field cooling (ZFC) and field cooling (FC) modes. (b) The high-temperature $\chi$($T$) data of sample $x=0.5$ under ZFC and FC modes. (c) Temperature dependence of the evaluated polarization $\Delta P$ and dielectric constant ($\varepsilon$). (d) $M-H$ loops measured at temperature $T=10$, $100$, $150$, and $300$ K for sample $x=0.5$. Reprinted from Ref.~\cite{Lin:Prb16}, with the permission of American Physical Society.}
\label{LSFO2}
\end{figure}

Magnetic susceptibility ($\chi$) for the $x=0.5$ sample does show a magnetic transition temperature $T_{\rm N}\sim167$ K at which ZFC and FC curves split, followed by a weak anomaly at $T_{\rm f}\sim162$ K at which ZFC curve peaks as shown in Fig.~\ref{LSFO2}(a). Usually, these two temperatures should be identical, indicating the magnetic transition. However, due to the intrisic quenching disorder caused by Sc substitution, there is a small difference (~5 K) in this system. Regarding the conflict of magnetism above room temperature in $h$-LuFeO$_3$ film, magnetic measurement on bulk samples shows a magnetic anomaly $\sim T_{\rm A}=445$ K as shown in Fig.~\ref{LSFO2}(b), which needs further study to figure out its origin. Magnetic hysterisis loop ($M-H$)(Fig.~\ref{LSFO2}(d)) suggests a weak ferromagnetic signal below $T_{\rm A}$, which is possibly the spin-canting moment from the AFM background as proposed by Wang \textit{et al.}. Once the AFM order is established, the weak ferromagnetic canting can be driven by the Dzyaloshinskii-Moriya interaction.

For the ferroelectric properties, (Lu$_{0.5}$Sc$_{0.5}$)FeO$_3$ are already ferroelectric at room temperature, which is driven by the freezing of the three collective phonon modes ($\Gamma_{2}^{-}$,$K_{1}$,$K_{3}$). Further dielectric constant shows a weak anomaly around $T_{\rm f}$ as shown in Fig.~\ref{LSFO2}(c). This additional ferroelectric transition was thought to originate from spin reorientation, which should be magnetoelectrically active. The saturated polarization can reach $\sim135$ $\mu$C/m$^2$ below $80$ K for polycrystalline samples. This direct pyroelectric polarization signal around magnetic transitions, which is a fingerprint of magnetoelectricity, is first observed among hexagonal $R$MnO$_3$ and $R$FeO$_3$ series.

\subsection{LiFe(WO$_4$)$_2$: cycloidal spins driving ferroelectricity}
As aforementioned, the weak magnetoelectric coupling in multiferroics with proper ferroelectricity is an intrinsic drawback, which is difficult to be overcome. Even in aforementioned geometric improper ferroelectrics, the independent origins of magnetism and ferroelectricity prohibit intrinsically strong magnetoelectricity. To realize strong magnetoelectric coupling and reliable magnetic/ferroelectric mutual regulation, one possible solution is the so-called type-II multiferroic family, in which their ferroelectricity is generated by some special spin orders. Since the ferroelectricity is a result of spin texture, changing the spin configurations (such as by applying magnetic field or other stimulations) can modulate ferroelectricity.

To obtain the type-II multiferroics, the special spin orders should break the spatial inversion symmetry, as required by the ferroelectricity. An effective approach is to stabilize cycloidal spin orders with helicity, i.e. clockwise and anti-clockwise. The first mostly studied system is orthorhombic TbMnO$_3$, in which the $bc$-plane spiral spin order forms below $28$ K. Such a spiral spin order generates a polarization pointing to the $c$-axis. This polarization can be switched from the $c$-axis to $a$-axis by magnetic field applied along the $b$-axis, rendering the strong magnetic-control of polarization \cite{Kimura:Nat,Arima:Prl}. The counter-effect, i.e. electric-control of magnetism, is more difficult to be observed, since here the magnetism is more fundamental than polarization in the type-II multiferroics. Even though, the tuning of magnetic helicity by an electric field has been observed \cite{Kagawa:Prl}, as an unambiguous evidence of electric-control of magnetism.

The underlying magnetoelectricity can be abstractly expressed as \cite{Sergienko:Prb,Katsura:Prl}:
\begin{equation}
\textbf{P}_{ij}\sim\textbf{e}_{ij}\times(\textbf{S}_{i}\times\textbf{S}_{j}),
\end{equation}
where $\textbf{S}$ denote spins and $\textbf{e}_{ij}$ is the direction vector from spin $\textbf{S}_{i}$ to spin $\textbf{S}_{j}$. The driving force for this type of magnetoelectricity is the Dzyaloshinskii-Moriya interaction. This equation is valid for lots of multiferroics in this catalog \cite{Kimura:Armr}.

In fact, a large portion of spiral-spin multiferroics are Mn oxides, e.g. $R$MnO$_3$ \cite{Goto:Prl,Dong:Prb08.2,Dong:Mplb}, $R$Mn$_2$O$_5$ \cite{Hur:Nat,Chapon:Prl,Lee:Prl13,Zhao:Sr}, MnWO$_4$ \cite{Taniguchi:Prl,Heyer:Jpcm}, and CaMn$_7$O$_{12}$ \cite{Zhang:Prb11,Johnson:Prl,Lu:Prl}. Besides these Mn oxides, others transition metal oxides are also available, as spiral-spin multiferroics, e.g. CuO \cite{Kimura:Nm}, CoCr$_2$O$_4$ \cite{Yamasaki:Prl06}, Ni$_3$V$_2$O$_8$ \cite{Lawes:Prl}. Among these materials, MnWO$_4$, is a tungstate member with the wolframite structure, which also displays frustrated magnetic orders. More than ten years ago, MnWO$_4$ was already experimentally confirmed to be a multiferroic material in the temperature range between $7.6$ K and $12.7$ K, corresponding to the incommensurate elliptical spiral phase \cite{Taniguchi:Prl,Heyer:Jpcm}. However, it is very strange that MnWO$_4$ is the sole multiferroic in the tungstate family for more than ten years, while in other families (e.g. $R$MnO$_3$) usually more than one multiferroic materials exist with similar magnetoelectric mechanism.

\begin{figure}
\centering
\includegraphics[width=0.96\textwidth]{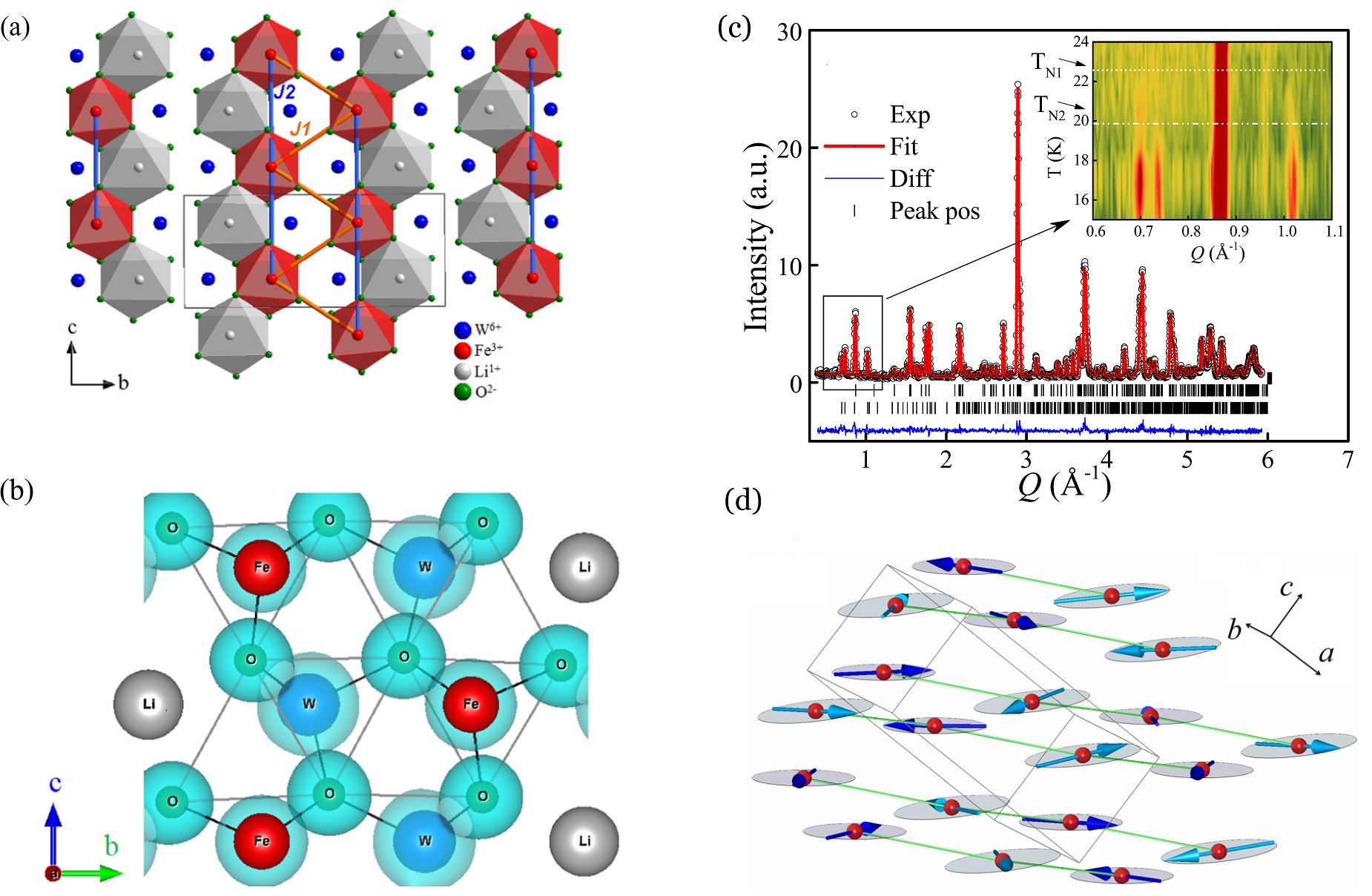}
\caption{(a) The crystal structure of LiFe(WO$_4$)$_2$, projected in the $bc$ plane. Blue: W; green: O; gray: Li; red: Fe. (b) Contour plot of charge density derived from density functional theory results with noncollinear magnetic state, $k=$($1$, $0$, $1/3$). The nearest neighbor Fe-O, W-O, and O-O are connected. (c) The neutron powder diffraction pattern measured at $5$ K and corresponding Rietveld fit. Inset: contour plot of the temperature dependence of magnetic Bragg peaks at small momentum transfer $\textbf{Q}$. (d)The sketch of noncollinear magnetic order fitted from the neutron powder diffraction data. The moments of Fe form a cycloidal structure with iron magnetic moments nearly confined to the plane defined by the $k$ vector and [$010$] direction. The cycloid rolls along the solid green lines. The moments at the two Fe positions related by the twofold axis symmetry, Fe1 ($0$, $y$, $1/4$), Fe2 ($0$, $-y$, $3/4$), are depicted by different colors. Reprinted from Ref.~\cite{LiuMf:PRB17}, with the permission from American Physical Society.}
\label{LFWO1}
\end{figure}

Very recently, some of the authors experimentally reported the second multiferroic tungstate LiFe(WO$_4$)$_2$, in which Li and Fe take place of Mn in MnWO$_4$ \cite{LiuMf:PRB17}. The crystal structure of LiFe(WO$_4$)$_2$ is described in the monoclinic space group $C2/c$. It consists of stacking (100) layers made of mixed [LiO$_6$] and [FeO$_6$] edge-sharing octahedra arranged in zigzag chains, separated by layers composed of tungstate [WO$_6$] octahedra. The chain contains both Li and Fe octahedra alternating along the $c$ direction. Such atomic arrangement leads to the doubling of the unit cell along the $b$ direction, thus it stands for a sub-branch of tungstate: double tungstate. Fig.~\ref{LFWO1}(a) shows the crystal structure of LiFe(WO$_4$)$_2$.

The spin order of this double tungstate was investigated by neutron and magnetic susceptibility measurements. Curie-Weiss fitting yields a negative $\theta_{\rm CW}=-69.5$ K, suggesting strong AFM interactions between Fe$^{3+}$ spins. An effective paramagnetic moment of $6.075$ $\mu_{\rm B}$ per Fe is found, which is very close to the expected value of effective moment ($5.92$ $\mu_{\rm B}$) for high-spin Fe$^{3+}$ ($S_{\rm z}=5/2$, $L=0$). Neutron diffraction results suggest a short range magnetic ordering at $22.6$ K and a long range spin order forms below $19.7$ K as shown in Fig.~\ref{LFWO1}(c). The refined magnetic structure of LiFe(WO$_4$)$_2$ is shown in Fig.~\ref{LFWO1}(d). The magnetic moments of Fe form a cycloidal magnetic structure with the spins confined to the plane defined by the $k$ vector and [$010$] direction. The envelope of the cycloid is nearly circular, with a refined amplitude of the magnetic moment of $4.2$ $\mu_{\rm B}$. The spin ordering temperature is remarkably enhanced with that in MnWO$_4$ ($12.7$ K). This enhancement is very precious, considering the fact that each [FeO$_{6}$]$'$s are separated by [WO$_{6}$]$'$s and [LiO$_{6}$]$'$s. Thus, the exchanges between Fe spins are not more indirect than those between Mn spins in Mn(WO$_4$)$_2$. Further density function theory calculations confirmed that the cycloidal magnetic structure as the possible ground state and revealed that the magnetic coupling between Fe$^{3+}$ ions can be mediated via Fe-O-O-Fe and Fe-O-W-O-Fe as shown in Fig.~\ref{LFWO1}(b). Such complicated exchange routes suppress the effective strength of magnetic couplings. Even though, thanks to the intrinsically stronger superexchange between Fe$^{3+}$-Fe$^{3+}$ than Mn$^{2+}$-Mn$^{2+}$, the magnetic N\'eel temperature remains improved.

\begin{figure}
\centering
\includegraphics[width=0.48\textwidth]{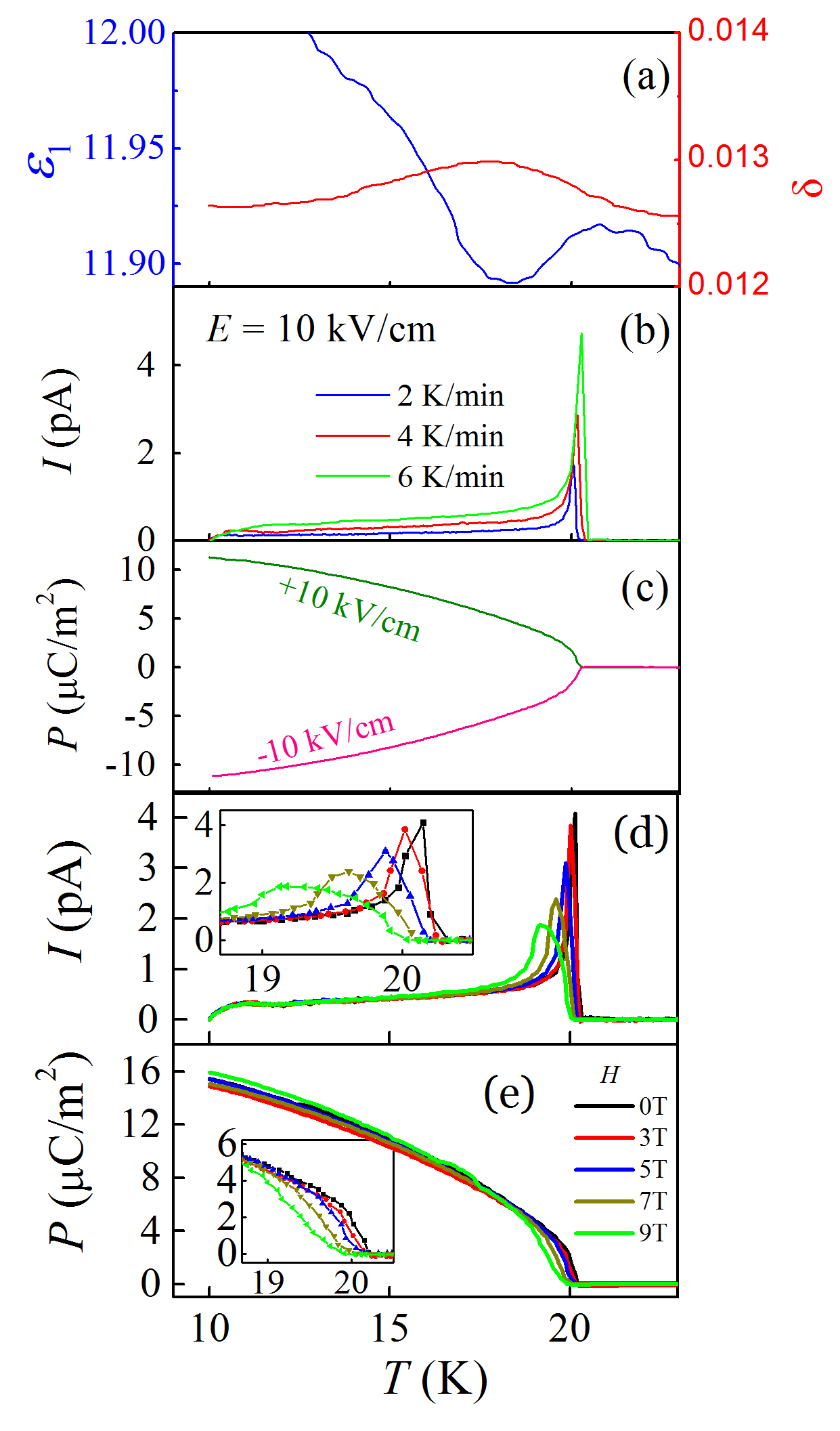}
\caption{(a) Dielectric constant (left) and dielectric loss (right). (b) Pyroelectric currents measured with different warming rates. The poling electric field is $10$ kV/cm. (c) Integrated pyroelectric polarization $P$'s with positive/negative poling fields. The peak position of dielectric constant coincides with the emergence of pyroelectric $P$'s. (d) Pyroelectric currents measured under different magnetic fields. (e) The corresponding pyroelectric $P$'s. Insets: magnified views around $T_{\rm N2}$. Reprinted from Ref.~\cite{LiuMf:PRB17}, with the permission from American Physical Society.}
\label{LFWO2}
\end{figure}

As mentioned before, this proposed cycloidal magnetic structure can break the spatial inversion symmetry and lead to the magnetic ferroelectricity. Dielectric constant $\varepsilon(T)$ measured at $1$ kHz [Fig.~\ref{LFWO2}(a), left axis] does show a broad peak around $T_{\rm N2}$, which is an indication of ferroelectricity below this temperature. Pyroelectric curves ($I_{\rm pyro}-T$) with three warming rates ($2$, $4$, and $6$ K/min) show peaks at $T_{\rm N2}$ without any shift (Fig.~\ref{LFWO2}(b)). Integrated polarization $P(T)$ curves based on positive and negative pooling electrical $I_{\rm pyro}-T$ curves are shown in Fig.~\ref{LFWO2}(c). The symmetrical $P(T)$ curves upon the positive/negative poling fields suggest the reversibility of polarization. According to $P(T)$ and $\varepsilon(T)$, the ferroelectricity emerges just below $T_{\rm N2}$. This is a strong evidence for magnetism driven ferroelectricity. Therefor, an intrinsic magnetoelectricity should be expected. $I_{\rm pyro}-T$ curves and corresponding $P(T)$ curves are measured under different magnetic fields as shown in Fig.~\ref{LFWO2}(d-e) with increasing magnetic field. Peaks at $T_{\rm N2}$ in both curves shifts to lower temperatures and becomes weaker and broader, implying the magnetoelectric coupling between magnetic order and dipole order. Density functional calculation also confirmed the cycloidal spin order driving ferroelectricity, and the estimated polarization was $24.5$ $\mu$C/m$^2$, in agreement with the experimental pyroelectric polarization ($\sim12$ $\mu$C/m$^2$ for polycrystalline sample) qualitatively.

Finally, it should be noted that there is a related compound NaFe(WO$_4$)$_2$, which also shows noncollinear spin order. However, it was found non-ferroelectric and its concrete positions of Na and Fe are different from Li and Fe in LiFe(WO$_4$)$_2$ \cite{Holbein:Prb2016}.

\subsection{Iron selenides: exchange striction driving ferroelectricity}
Besides cycloidal spin order, some collinear spin order, e.g. $++--$ type, can also break spatial reversal symmetry for particular crystalline structures. The first material in this catalog was orthorhombic HoMnO$_3$/YMnO$_3$ as proposed by Sergienko, \c{S}en and Dagotto, which owns the zigzag E-type antiferromagnetism \cite{Sergienko:Prl}. Another early member is Ca$_3$CoMnO$_6$ with quasi-one-dimensional magnetic chains \cite{Choi:Prl}. Although these materials also belong to type-II multiferroics, the underlying magnetoelectric mechanism is rather different from the above cycloidal spin driven one. Generally speaking, the cycloidal ones need the relativistic spin-orbit coupling to generate the polarizations, and thus the polarizations are usually very weak considering the strength of spin-orbit coupling of most transition metal ions. In contrast, the driving force in HoMnO$_3$/YMnO$_3$ and Ca$_3$CoMnO$_6$ is the so-called exchange striction, which can be abstractly described as $J \textbf{S}_{i}\cdot\textbf{S}_{j}$, where $J$ is the exchange coefficient. Since in most materials the exchange is much stronger than the Dzyaloshinskii-Moriya interaction, the induced polarization can be much larger than the cycloidal spin order induced one, e.g. up to $\sim1$ $\mu$C/cm$^2$ in orthorhombic YMnO$_3$ \cite{Sergienko:Prl,Nakamura:Apl}. However, the multiferroic temperatures remain low in most materials in this catalog.

\subsubsection{BaFe$_2$Se$_3$}
This exchange striction mechanism can also work in non-oxides. For example,  iron-selenide BaFe$_2$Se$_3$ was predicted by some of the authors to be a type-II multiferroic, whose polarization is driven by exchange striction \cite{Dong:PRL14}. This work, not only finds a new multiferroic material, but also connect the multiferroics with superconductor family, since BaFe$_2$Se$_3$ is a member of iron-based superconductor family.

\begin{figure}
\centering
\includegraphics[width=0.96\textwidth]{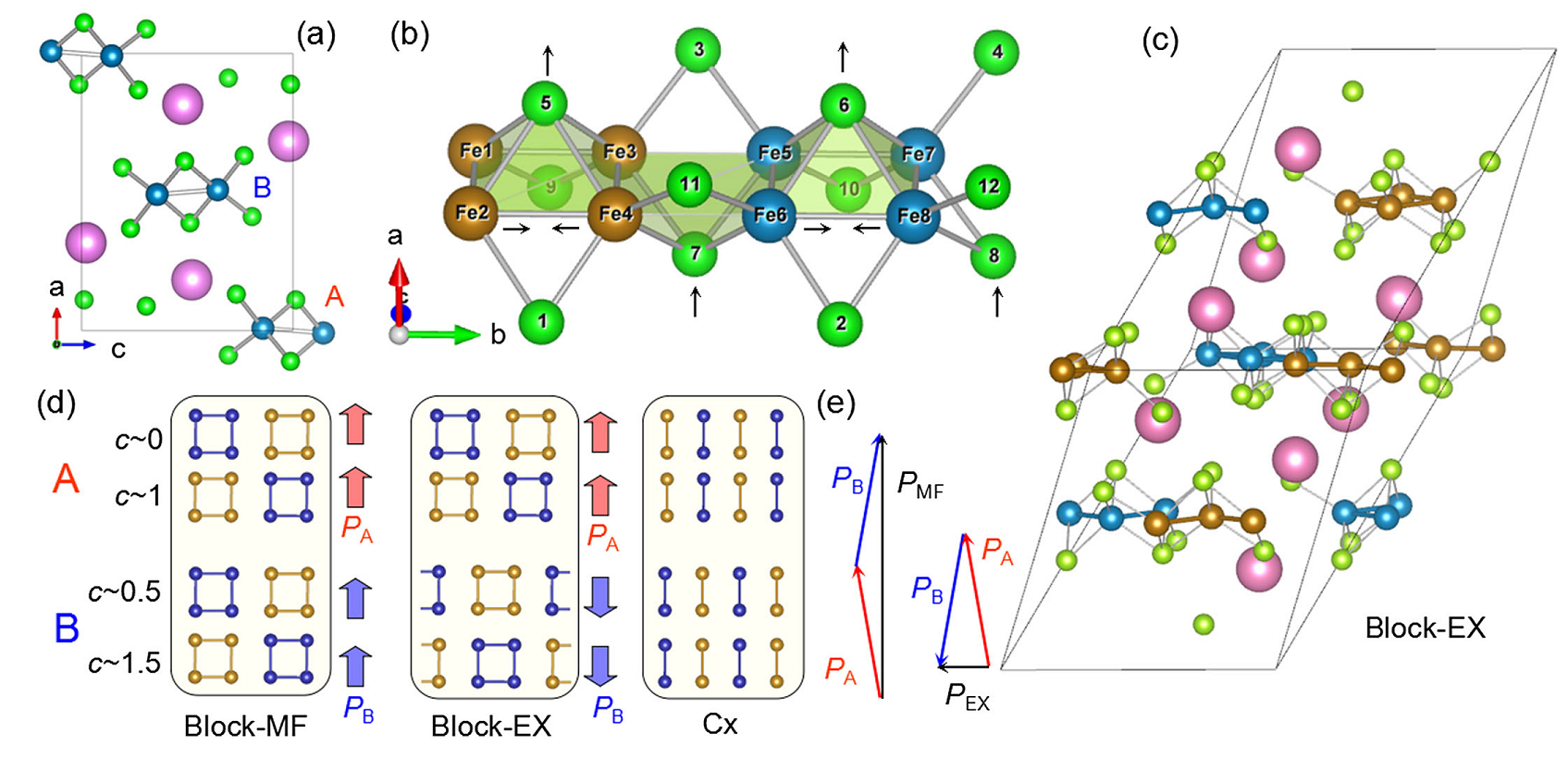}
\caption{Crystal and magnetic structures of BaFe$_2$Se$_3$. (a) Side view along the $b$ axis. Blue: Fe; green: Se; pink: Ba. (b) A Fe-Se ladder along the $b$ axis and its magnetic order. Partial ionic displacements driven by the exchange striction are marked as black arrows. (c) A unit cell considering the AFM order. (d) Spin structures. Left: Block-MF (MF: multiferroic); middle: Block-EX (EX: experimental); right: Cx (C-type stripe along the $x$ axis). The side arrows denote the local ferroelectric polarization of each ladder. In (b)-(d), the spins ($\uparrow$/$\downarrow$) of Fe's are distinguished by colors. (e) Vector addition of ferroelectric polarization's of ladders A and B. Reprinted from \cite{Dong:PRL14}, with the permission from American Physical Society.}
\label{BFS}
\end{figure}

BaFe$_2$Se$_3$ forms an orthorhombic structure. Each unit cell has two iron ladders (labeled A and B), built by edge-sharing FeSe$_4$ tetrahedra, as shown in Fig.~\ref{BFS}(a). Long-range AFM order is established below $256$ K. Both neutron studies and first principles calculations reported an exotic block AFM order \cite{Caron:Prb12,Nambu:Prb,Saparov:Prb,Medvedev:Jl}. The Hartree-Fock approximation to the five-orbital Hubbard model also confirmed the stability of the block AFM phase and revealed other competing phases, e.g. the Cx phase \cite{luo:prb13}.

The block AFM order is particularly interesting because it breaks parity symmetry and displays exchange striction effects. Neutron measurements revealed that the nearest-neighbor (NN) distances between Fe($\uparrow$) and Fe($\uparrow$) [or Fe($\downarrow$) and Fe($\downarrow$)] at $200$ K become $2.593$ \AA, much shorter than the Fe($\uparrow$) $-$ Fe($\downarrow$) distance $2.840$ \AA{} \cite{Caron:Prb12}. However, such in-ladder striction between irons will not generate polarization. The polarization comes from the displacements of Se's.

As shown in Fig.~\ref{BFS}(b), Se(5) is above the ladder's plane, while the next Se(7) is below, and the distances of Se(5) and Se(7) to the iron ladder plane should be the same in magnitude and opposite sign (``antisymmetric"). However, this antisymmetric could be broken by the block AFM order. The blocks made of four Fe($\uparrow$)'s [or four Fe($\downarrow$)'s] are no longer identical to blocks made of two Fe($\uparrow$)'s and two Fe($\downarrow$)'s \cite{Dong:PRL14,Zhang:Prb2018}. Then, the Se(5) and Se(7) heights do not need to be antisymmetric anymore; their distances to the ladder planes can become different. The same mechanism works for the edge Se's, e.g. Se(1) and Se(11). As a consequence, the atomic positions of Se break the space inversion symmetry, generating a local polarization pointing perpendicular to the iron ladders plane (almost along the $a$ axis).

Theoretical analysis suggests that each ladder can be multiferroic. However, the net polarization is determined by how each ladder interact with others. Neutron studies by Caron \textit{et al.} found the block AFM pattern shows a $\pi/2$-phase shift between the NN A-B ladders but a $\pi$-phase shift between the NN A-A ladders (and NN B-B ladders), i.e. Block-EX state as shown in Fig.~\ref{BFS}(d)\cite{Caron:Prb,Caron:Prb12}. The дл shift between A-A ladders (or B-B ladders) will not change the direction of the induced local polarization, but the $\pi/2$-phase shift between A-B ladders will induce (nearly) opposite local polarization. A fully cancellation can be avoided due to the small canting angles between ladder A and B, leading to a net polarization along $c$ axis as shown in Fig.~\ref{BFS}(e).

According to the density functional theory calculation, the Block-EX state are found to be ferrielectric with a net polarization $\sim$ $0.19$ $\mu$C/cm$^2$ pointing mostly along $c$ axis. The net polarization can be flipped if an external electric field is applied along $c$ axis. If a large enough field is applied along the $a$ axis, the ferrielectric (Block-EX) to FE (Block-MF) phase transition will occur, producing a $90^\circ$ flipping and enhancement of polarization. The ferroelectric polarization of Block-MF state can reach $2$ $\mu$C/cm$^2$. Moreover, the $180^\circ$ flipping of polarization can also be obtained by reversing the tilting angle between the planes of ladders A and B, without shifting the magnetic blocks \cite{Dong:PRL14}.

However, due to the nonstoichiometry of BaFe$_2$Se$_3$ and thus low resistivity, the direct measurement of polarization becomes difficult. Even though, following neutron scattering already confirmed the polar structure \cite{Lovesey:PS}. And using high-resolution transimission electron microscope, Tian \textit{et al.} observed the local dipoles of each iron ladder (private communication).

\subsubsection{KFe$_2$Se$_2$}
\begin{figure}
\centering
\includegraphics[width=0.96\textwidth]{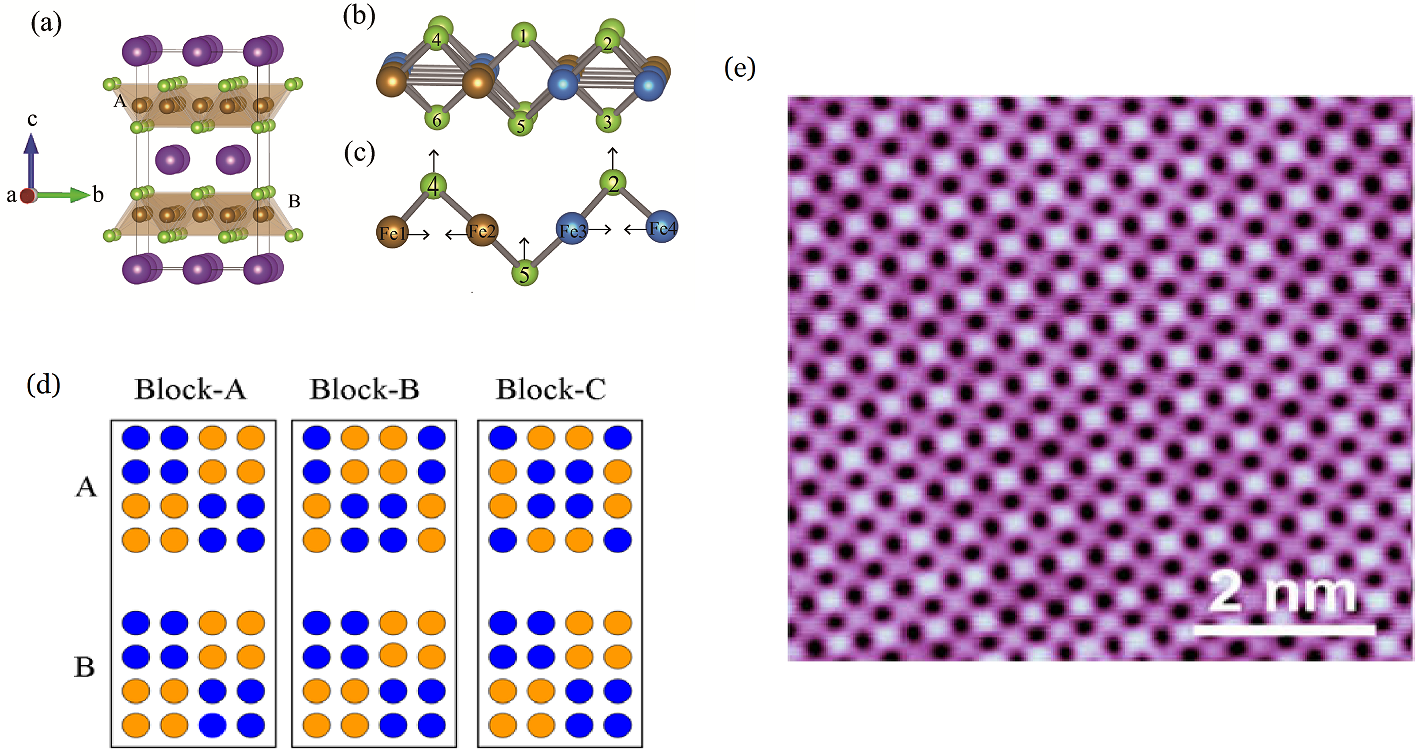}
\caption{(a) Crystal structure of KFe$_2$Se$_2$. Purple: K; green: Se; brown: Fe. Two Fe sheets in a minimum unit cell are indicated as A and B. (b) One FeSe layer with magnetism. Brown: spin up; blue: spin down. (c) A side view of FeSe bonds. The ionic displacements driven by exchange striction are indicated by arrows. (d) Sketch of the block AFM series. A and B denote the two layers shown in (a). Left: Block-A. Middle: Block-B. Right: Block-C. Irons with spin up and spin down are distinguished by colors. Reprinted from \cite{Zhang:rrl}, with the permission from John Wiley\&Sons, Inc.(e) The charge ordering in STM image of stoichiometric KFe$_2$Se$_2$. Reprinted from \cite{Li:Prl}, with the permission from American Physical Society.}
\label{KFS}
\end{figure}

Besides the $123$-type iron-selenide BaFe$_2$Se$_3$, stoichiometric KFe$_2$Se$_2$ also show the block-type antiferromagnetism, which is not common in iron-based superconductors. KFe$_2$Se$_2$ forms the tetragonal crystal structure, whose space group is $I4/mmm$ (No. $139$). In each unit cell, there are two Fe layers, each of which is built by edge-sharing FeSe$_4$ tetrahedra. K ions intercalate between Fe$-$Se layers as shown in Fig.~\ref{KFS}(a).

The block-type antiferromagnetism was first predicted by Li \textit{et al.} according to the density functional theory calculation \cite{Li:Prb12}, then its associated structural tetramerization was experimental revealed in KFe$_2$Se$_2$ thin film using scanning tunnelling microscope (STM) measurement by Xue's group \cite{Li:Np,Li:Prl}(Fig.~\ref{KFS}(e)). The tetramerization of irons is driven by its block-AFM ordering. The NN distance between Fe($\uparrow$) and Fe($\uparrow$) [or Fe($\downarrow$)-Fe($\downarrow$)] is shorten comparing with the one between Fe($\uparrow$)-Fe($\downarrow$), similar to the situation in BaFe$_2$Se$_3$. Se(4) and Se(5) are located in the opposite sides of iron layer, as shown in Fig.~\ref{KFS}(b). Originally, the distance of Se(4) and Se(5) to the iron layer should be identical. Due to the tetramerization, the shrunk Fe($\uparrow$) blocks will push Se(4) ion upward, while Se(5) will be lifted up due to the elongation of Fe($\uparrow$)$-$Fe($\downarrow$) bond as shown in Fig.~\ref{KFS}(c). Similar movements occur for other Se ions. As a result, the movements of Se ions break the inversion symmetry and generate a local dipole moment pointing perpendicular to the iron plane, i.e. along the $c$-axis \cite{Zhang:rrl}.

Each Fe$-$Se layer should be multiferroic, this layered system can be either a ferroelectric one with a finite macroscopic ferroelectric polarization or an antiferroelectric one with canceled polarization, depending on the stacking of magnetic blocks along the $c$-axis. As shown in Fig.~\ref{KFS}(d), there are three possible block-AFM order. The magnetic + lattice space group changes from No. $139$ ($I4/mmm$) to No. $51$ ($Pmma$ for Block-A) or No. $36$ ($Cmc2_1$ for Block-B) or No. $123$ ($P4/mmm$ for Block-C), among which only the $Cmc2_1$ for Block-B is a polar space group. Density functional theory calculation performed by Zhang \textit{et al.} observed the difference in the bond lengths between Fe($\uparrow$)-Fe($\uparrow$) [or Fe($\downarrow$)$-$Fe($\downarrow$)] and Fe($\uparrow$)-Fe($\downarrow$), e.g. $2.542$\AA and $2.929$ \AA, respectively\cite{Zhang:rrl}. Due to exchange striction, the heights of Se (to the iron plane) become different: $1.53$ \AA{} for Se(4) and $1.58$ \AA{} for Se(5), respectively. The net polarization calculated using the standard Berry phase method for block B is along the $c$-axis with magnitude increases from $0.48$-$2.1$ $\mu$C/cm$^2$ depending on the choice of $U$ in calculation \cite{Zhang:rrl}. For Block-A and Block-C, the dipole moments between any nearest-neighbor layers are aligned antiparallelly, rendering the antiferroelectric fact.

According to the calculation, the block AFM series always own lower energies than other type of magnetic order, such as ferromagnetic, A-, C-, or G-type AFM states. Among the three types of block AFM, the ferroelectric Block-B owns the lowest energy at small effective $U$, while the antiferroelectric Block-C take places with increasing effective $U$. More importantly, the energy differences within the block AFM series are very tiny. Therefore, antiferroelectric-ferroelectric transition is possible via proper stimulates in this material.

\subsubsection{CaOFeS}
\begin{figure}
\centering
\includegraphics[width=0.96\textwidth]{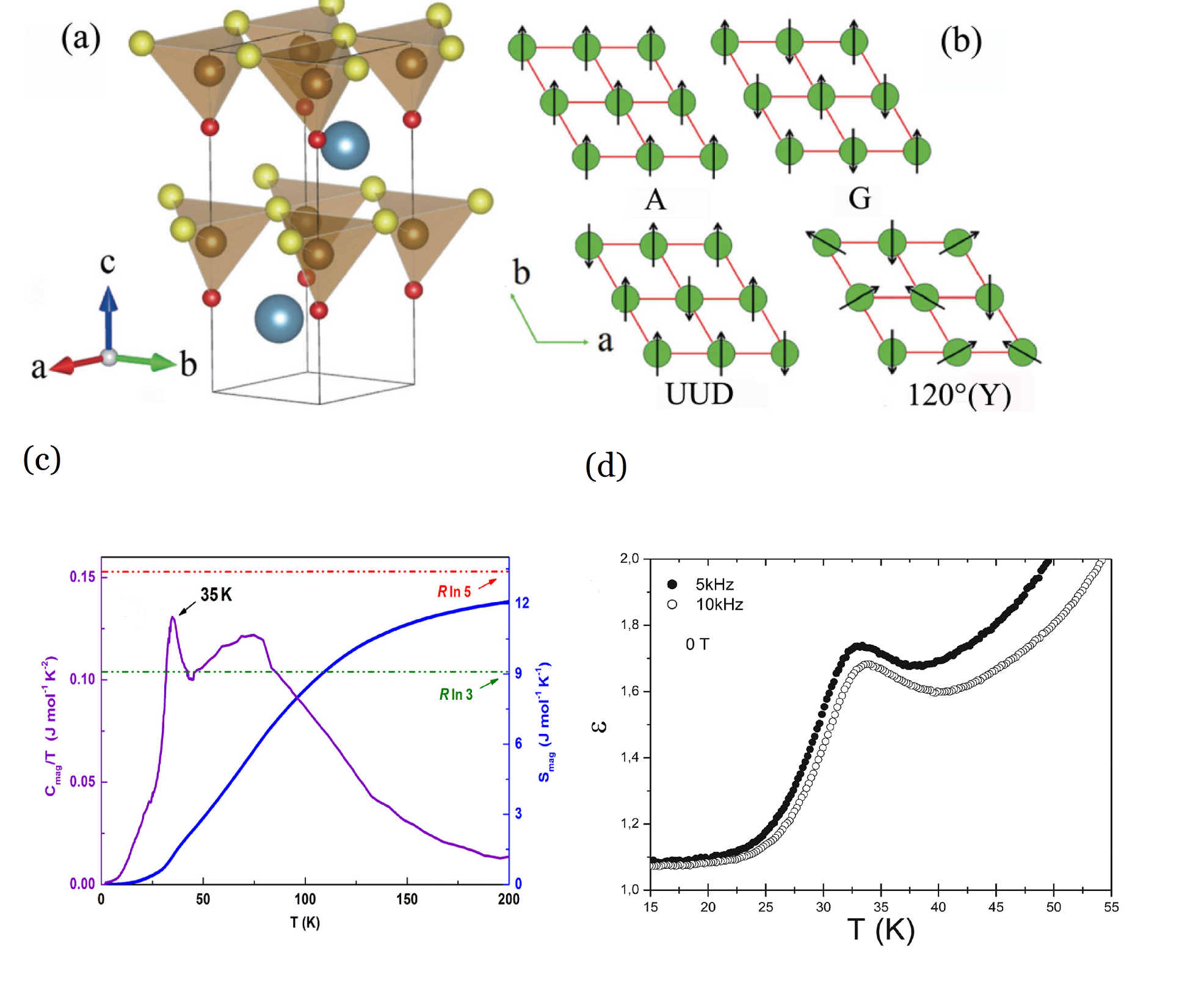}
\caption{(a) Schematic crystal structure of CaOFeS. Blue:Ca;red: O; yellow: S; brown: Fe. (b) Sketch of possible spin configurations (denoted by arrows) in a two-dimensional triangular lattice. Between layers, both the parallel and antiparallel configurations have been calculated. Reprinted from \cite{zhang:prm}, with the permission from American Physical Society. (c) Temperature dependence of magnetic entropy and $C_{\rm mag}/T$ for CaOFeS. Reprinted from \cite{Jin:prb}, with the permission from American Physical Society. (d) Thermal dependence of the dielectric permittivity $\varepsilon$ (measuring frequency $f=5$ and $10$ kHz). Reprinted from \cite{Delacotte:IC}, with the permission from ACS publication.}
 \label{COFS1}
\end{figure}

Besides aforementioned $123$ and $122$ iron selenides, the so-called $1111$ series of iron oxysulfides can also be magnetoelectric. Usually, the 1111 series of iron pnictides, e.g. LaOFeAs, own a layered Fe square lattice, which undergoes a tetragonal-to-orthorhombic structural transition followed by the stripe AFM transition \cite{Dai:Rmp,Cruz:Nat}. However, here the $1111$-type iron oxysulfide CaOFeS forms a layered triangular lattice \cite{Selivanov:IM,Sambrook:IC,Jin:prb,Delacotte:IC}. As sketched in Fig.~\ref{COFS1}(a), it owns a hexagonal structure, whose space group is $P6_3mc$ (No. $186$). In each unit cell, there are two $ab$-plane Fe layers, which are built by triangles of O-Fe-S$_3$ tetrahedra. Ca ions intercalate between S and O layers. This triangular lattice may provide the geometry for magnetic frustration. As shown in Fig.~\ref{COFS1}(b), if the spin is for Heisenberg type, a typical Y-type ground state usually appears with nearest-neighbor spins arranged with $120^\circ$ in the two-dimensional triangular lattice. While if the spin is for Ising type, spins arranged in a two-dimensional triangular lattice can also form some exotic patterns. Neutron measurement performed by Jin \textit{et al.}found a partially ordered G-type Ising type AFM with a propagation vector of $k=$($1/2$, $1/2$, $0$) and an ordered magnetic moment of $2.59(3)$ $\mu_{\rm B}$/Fe along $c$ at $6$ K \cite{Jin:prb}.

Dielectric measurements performed by Delacotte \textit{et al.} revealed the existence of a magnetodielectric effect near $33$ K as shown in Fig.~\ref{COFS1}(d), which is in good agreement with the N\'eel temperature $35$ K (Fig.~\ref{COFS1}(c)) \cite{Delacotte:IC}. Calculation performed by Zhang \textit{et al.} confirmed the G-type AFM (G-AFM) ground state and explained the mechanism of this magnetodielectric effect \cite{zhang:prm}.

\begin{figure}
\centering
\includegraphics[width=0.48\textwidth]{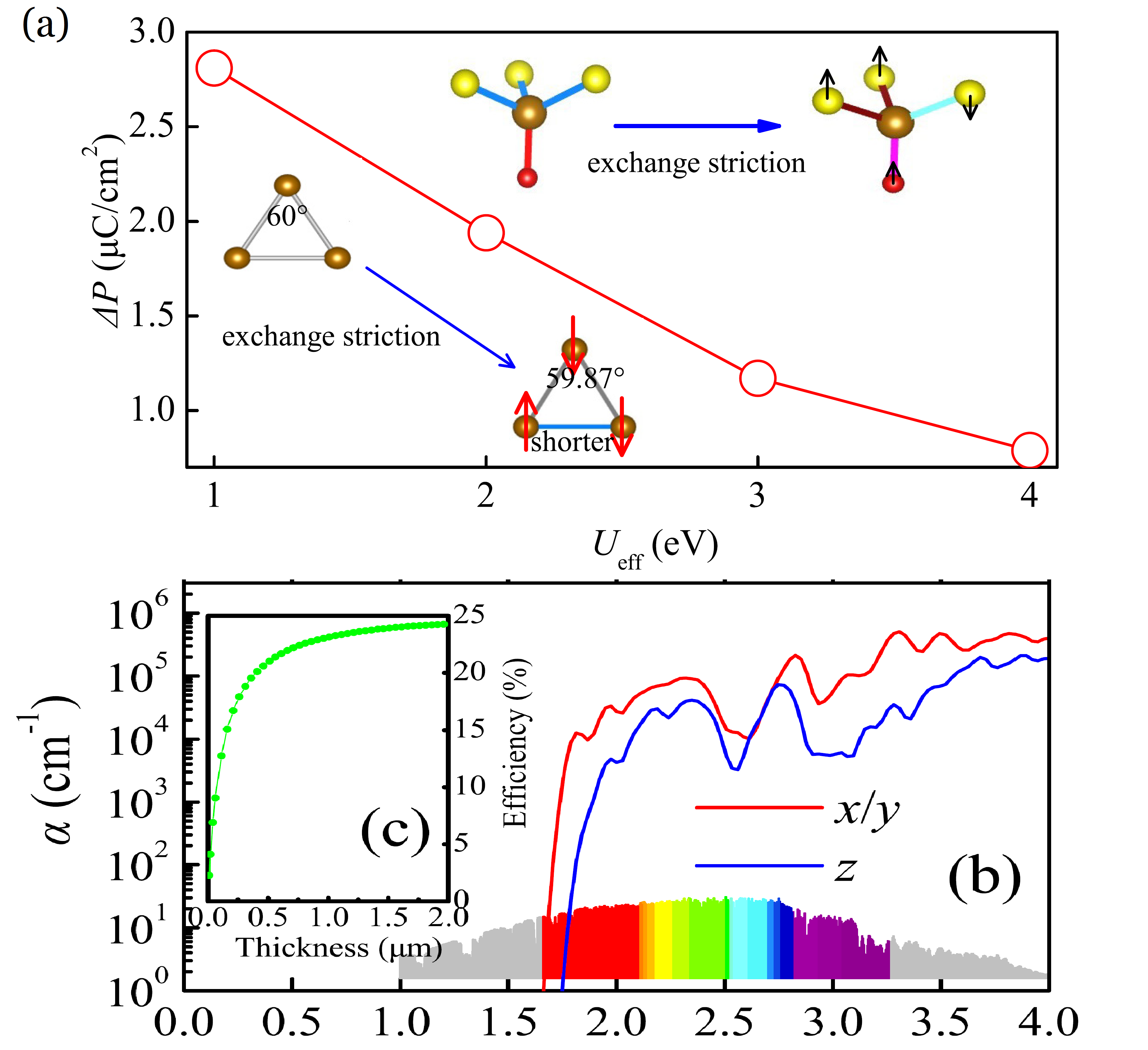}
\caption{(a) The magnetoelectricity, i.e. change of polarization upon G-AFM ordering. (b) The calculated absorption coefficient $\alpha(\omega)$ of CaOFeS. The energy spectrum of solar light is shown for reference. (c) Calculated maximum photovoltaic energy conversion efficiency for CaOFeS as a function of absorber layer thickness. Reprinted from \cite{zhang:prm}, with the permission from American Physical Society.}
\label{COFS2}
\end{figure}

The crystalline structure of CaOFeS, with a space group $P6_3mc$ and point group $6mm$, is polar, due to the unequivalence of O and S. But this polar structure is irreversible since the layers of O and S are fixed. The special G-AFM order breaks the trigonal (i.e. $120^\circ$ rotation) symmetry of the triangular lattice. In each Fe triangle, there are one Fe($\uparrow$)-Fe($\uparrow$) [or Fe($\downarrow$)-Fe($\downarrow$)] bond and two Fe($\uparrow$)-Fe($\downarrow$) bonds, which are no longer symmetric. This breaking of symmetry will distort the lattice, by shrinking the Fe($\uparrow$)-Fe($\downarrow$) bonds but elongating others. According to the density functional theory optimized structure, such exchange striction will also result in the change of Fe-Fe distance up to $0.008$ \AA, i.e. the triangles are no longer regular but with $0.13$ \AA{} correction for $\angle_{\rm Fe-Fe-Fe}$ as shown in Fig.~\ref{COFS2}(a) \cite{zhang:prm}. Such a tiny distortion is beyond the current experimental precision of structural measurement. The distortion of Fe-S bonds are more serious, reaching $0.069$ \AA{} as mentioned before. It is the displacements of S ions along the $c$ axis responsible for the observed magnetodielectric effect.

Strickly speaking, CaOFeS review here is not a multiferroic since it is polar but not ferroelectric. However, it shows the magnetoelectric effect and underlying mechanism is common with some type-II multiferroics with exchanges striction.

In addition, theoretic study predict a large coefficient of visible light absorption shown in Fig.~\ref{COFS2}(b). The maximum photovoltaic energy conversion is estimated to be $\sim24.2\%$ \cite{zhang:prm}. Compared with the estimated efficiency of some other photovoltaic materials, e.g. AgInTe$_2$ ($27.6\%$), CuBiS$_2$ ($16\%$), CH$_3$NH$_3$PbI$_3$ ($30\%$), and CuBiS$_2$ ($22\%$) \cite{Yu:prl12,Yu:AEM,Yin:AM14}, this efficiency is still valuable. Considering that the polar effect will enhances the electron-hole separation which has not been take into account in this estimation, CaOFeS may be a potential photovoltaic material with prominent efficiency.

\subsection{LiFe$_2$F$_6$: charge ordering driving ferroelectricity}
The type-II multiferroics as we discussed in Sections $2.2$ and $2.3$ have strong magnetoelectric coupling since their ferroelectricity directly originates from magnetism. However, in these materials, the magnetism $M$ is a primary parameter, but the polarization $P$ is not. Thus it is easy to control $P$ via magnetic field, but it is not easy to obtain the counter-effect. To overcome this drawback, the improper electronic ferroelectricity (or so-called charge ordering driving ferroelectricity) may provide a solution.

Iron-based compounds are naturally candidate for charge ordering type multiferroics since the multiple stable valence state of iron. For example, the first proposed multiferroics material generated by charge ordering, LuFe$_2$O$_4$, its ferroelectric is reported to be induced by a combination of the bilayer character of the crystal structure and the frustrated charge ordering in each layer \cite{Ikeda:Nat,subramanian:AM06,Nagano:PRL07,Xiang:PRL07,Zhang:PRL07}. Another example is Fe$_3$O$_4$. Fe$_3$O$_4$ becomes  ferrimagnetic below $860$ K, following by the famous Verwey transition at $120$ K. It displays ferroelectricity below the Verwey temperature due to the charge ordering between Fe$^{3+}$ and Fe$^{2+}$ ions on B site that surrounded by oxygen octahedron \cite{Alexe:Am}. However, there are drawbacks for both materials. The ferroelectricity mechanism of LuFe$_2$O$_4$ was questioned recently \cite{Groot:Prl,Niermann:Prl,Angst:RRL13}. Fe$_3$O$_4$ is a narrow band insulator below Verwey temperature, resulting in leaking current in experimental demonstration \cite{Fonin:prb05,liu:NPJQM16}.

Recently, an iron-based fluoride LiFe$_2$F$_6$ was predicted to be a rare multiferroic with both large magnetization and polarization mediated by charge ordering \cite{Lin:prm17}. The stronger electronegativity of F may reduce the hybridization between the $2p$ bands and $3d$ bands, leading to more insulating materials.	

\begin{figure}
\centering
\includegraphics[width=0.48\textwidth]{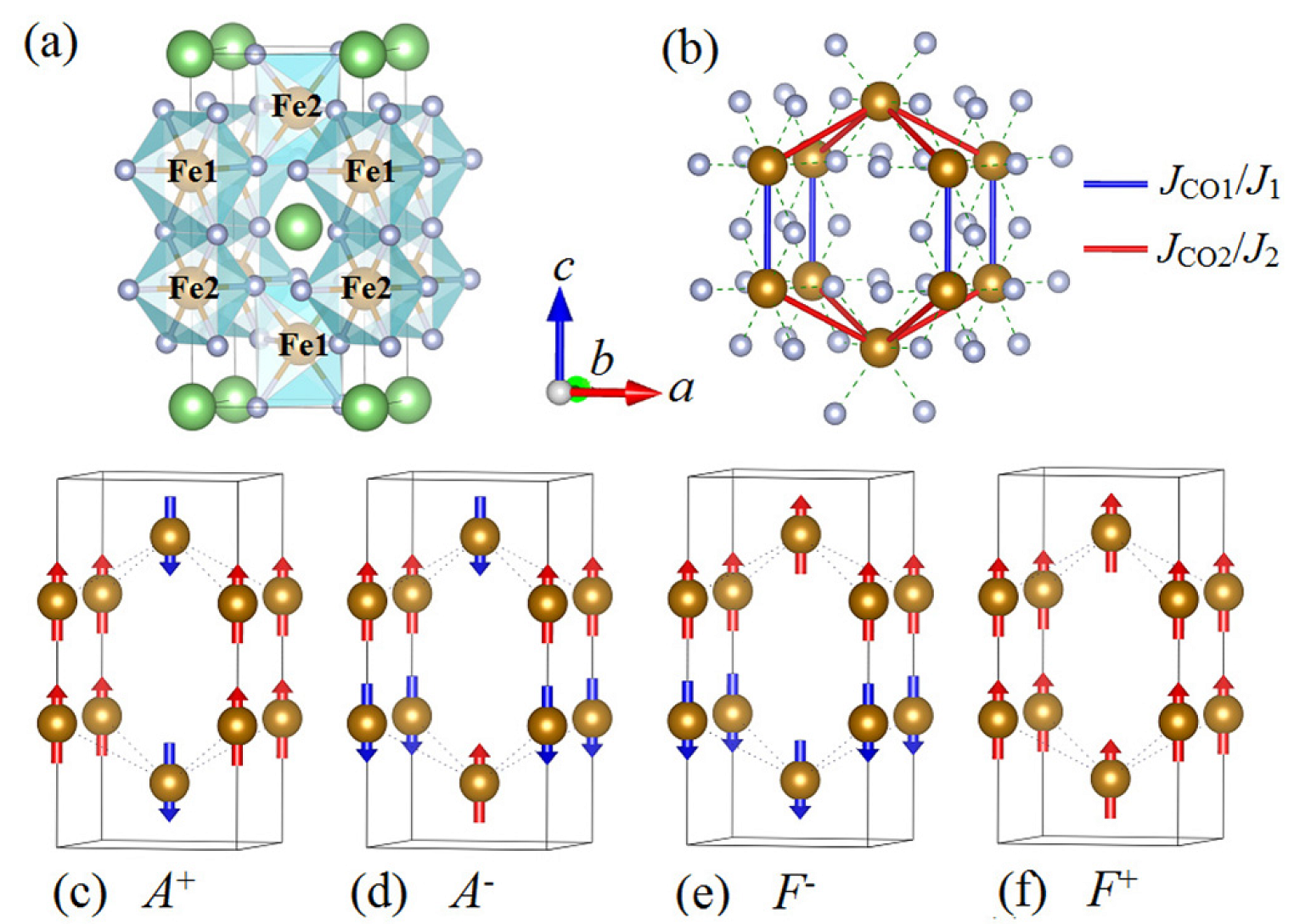}
\caption{(a) Crystal structure of LiFe$_2$F$_6$. Brown: Fe; green: Li; silver: F. HSS: Fe1=Fe2. LSS: Fe$\neq$Fe2. (b) The framework of Fe-F ions and the charge/magnetic exchange paths for $J_{\rm CO1}$/$J_1$ and $J_{\rm CO2}$/$J_2$. (c-f) Sketch of different magnetic orders of Fe spins. F and A ($+$ and $-$) stand for ferromagnetic and AFM coupling between next-nearest neighbor (nearest neighbor) moments, respectively. Reprinted from \cite{Lin:prm17}, with the permission from American Physical Society.}
\label{LFF1}
\end{figure}

LiFe$_2$F$_6$ forms a tetragonal crystal structure (Fig.~\ref{LFF1}(a)). At high temperature, the high symmetric structure (HSS) without charge ordering is $P4_{2}/mnm$ (No. $136$). M\"ossbauer spectrum measurement found the existence of Fe$^{2+}$ and Fe$^{3+}$ in LiFe$_2$F$_6$ above room temperature \cite{Greenwood:JCSA71,Fourquet:JSSC88}. Later, Fourquet \textit{et al.} studied a LiFe$_2$F$_6$ single crystal using x-ray diffraction\cite{Fourquet:JSSC88}. It revealed a low symmetric structure (LSS, No. $102$, $P4_{2}nm$) for the charge ordering state. Even the charge ordering temperature is not fully determined, it is definitely above room temperature. Further, Neutron powder diffraction revealed an A+ antiferromagnetism (Fig.~\ref{LFF1}(c)) below $105$ K \cite{Shachar:prb72,Wintenberger:SSC72}.

\begin{figure}
\centering
\includegraphics[width=0.48\textwidth]{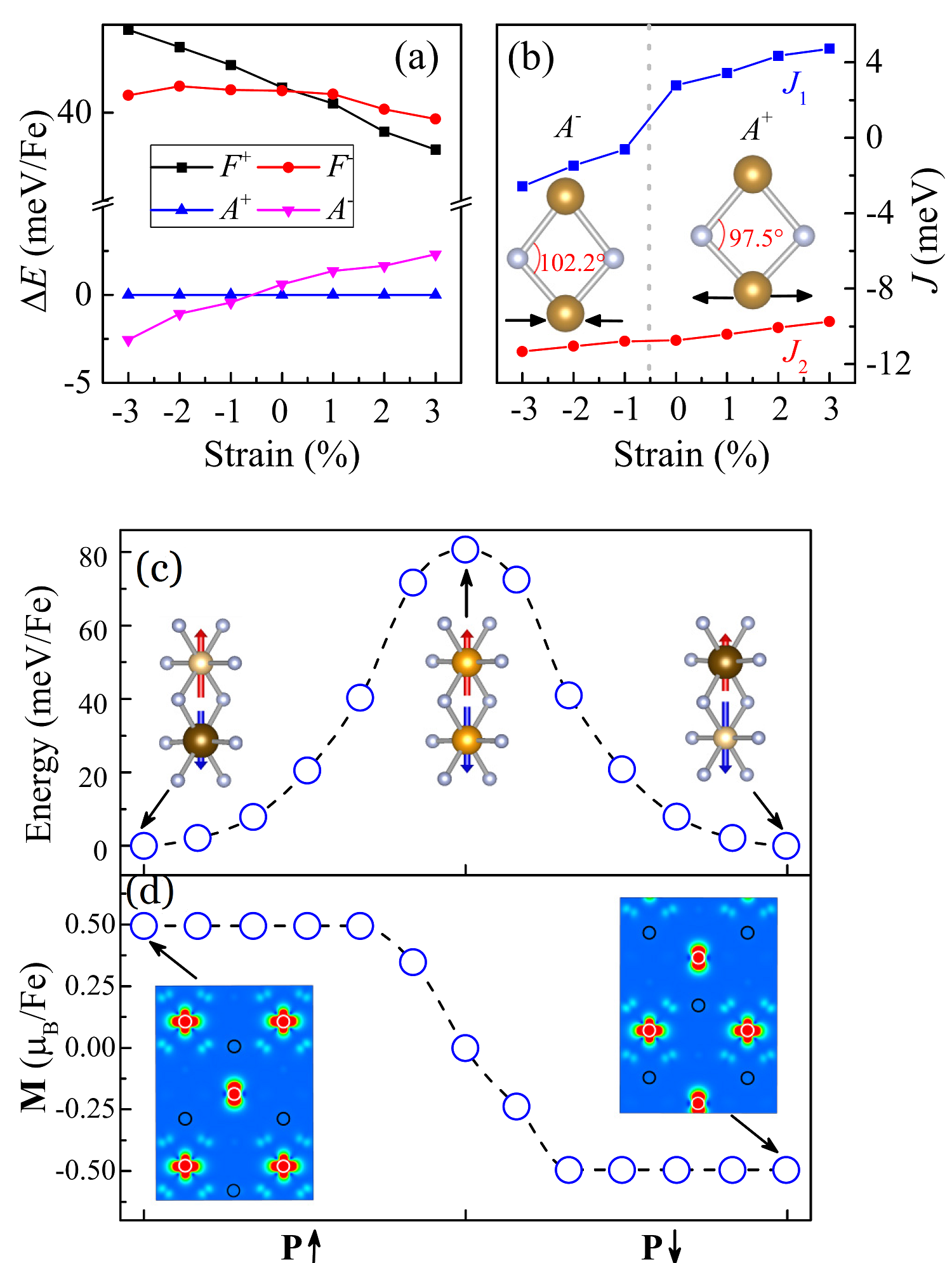}
\caption{(a) Results for strained LiFe$_2$F$_6$. (a) Energies of the F$^+$, F$^-$, A$^+$, and A$^-$ states as a function of strain. The energy of A$^+$ state is the reference. (b) Strain dependent exchanges for the Heisenberg model. Insert: sketch of the bond angle of Fe$^{2+}$-F-Fe$^{3+}$ for the $J_1$ path. (c) Sketch of CO-mediated magnetoelectricity in strained ($-3\%$) LiFe$_2$F$_6$. (a) Switch of ferroelectric $\textbf{P}$ simulated by the NEB method. The simulated energy barrier for switching should be considered as the upper limit in the real experiment, while other paths/processes with lower barriers are possible. Insets: Initial, intermediate, and final structures. (b) The corresponding switch of magnetic $\textbf{M}$ obtained in the NEB process. Insets: The corresponding profiles (viewed from the [110] direction) of the partial charge density for the topmost valence band. White/black circle: Fe$^{2+}$/Fe$^{3+}$. Reprinted from \cite{Lin:prm17}, with the permission from American Physical Society.}
\label{LFF2}
\end{figure}	

On the basis of these experimental observation, Lin \textit{et al.} performed theoretical study on the ferroelectric properties of this compound \cite{Lin:prm17}. They demonstrate that LiFe$_2$F$_6$ is an AFM ferroelectric. The ferroelectricity are indeed induced by charge ordering. Thus, the charge order transition is also the nonpolar/polar transition. The dipole moment formed by the Fe$^{2+}$-Fe$^{3+}$ pair is estimated to be $12.4$ $\mu$C/cm$^2$. More interestingly, calculation found that the energy of A-AFM state as shown in Fig.~\ref{LFF1}(d) with a net magnetization $0.5$ $\mu_{\rm B}$/Fe is only slightly higher ($0.5$ meV/Fe) than that of the ground state A$^+$. It can be achieved via compressive strain beyond $-0.5\%$. It is expected to flip the net magnetization together with the polarization by an electric voltage as shown in Fig.~\ref{LFF2}(a) and (b), which avoids the drawback of magnetic ferroelectrics, provides the desired magnetoelectric function in practice.

\section{Summary \& Perspective}
\begin{table}
\caption{Comparison of four improper ferroelectrics and their basic physical characteristics. ME: magnetoelectric.}
\begin{tabular*}{\textwidth}{@{\extracolsep{\fill}}lcccc}
\hline \hline
& geometric & cycloidal spin & exchange striction & charge ordering\\
Material & LuFeO$_3$ & LiFe(WO$_4$)$_2$ & BaFe$_2$Se$_3$ & LiFe$_2$F$_6$\\
Ferroelectric $T_{\rm C}$ & $\sim1050$ K & $\sim20$ K & $\sim256$ K & $\sim400$ K\\
Magnetic $T_{\rm N}$ & $\sim400$/$160$ K & $\sim23$ K & $\sim256$ K & $\sim100$ K\\
ME Coupling  & spin-lattice & spin-orbit & spin-lattice & spin-charge\\
\hline \hline
\end{tabular*}
\label{table1}
\end{table}

In this topical review, the concept of improper ferroelectrics was introduced. Its mechanisms and advantages compared to proper ferroelectrics were also discussed. It is noticeable that the focused systems of improper ferroelectrics have shifted gradually from Mn-based compounds to iron-based compounds. Recent progresses on some typical iron-based improper ferroelectrics have been reviewed in this article, which covered all three types of improper ferroelectrics. Several typical materials are reviewed, which are briefly summarized in Table~\ref{table1}. The first example is the hexagonal (Lu$_{0.5}$Sc$_{0.5}$)FeO$_3$, which is a geometric ferroelectric material. Theoretical calculation confirms that the partial substitution of Lu by Sc can stabilize the hexagonal structure while its improper ferroelectricity is not affected. Direct pyroelectric polarization signal around the low temperature magnetic transitions is observed as a fingerprint of magnetoelectricity. The second example is LiFe(WO$_4$)$_2$, which is found to be the second multiferroic tungstate. Although the effective strength of magnetic coupling is suppressed by complicated indirect exchange routes, the magnetic N\'eel temperature remains improved compare with the first reported multiferroic tungstate Mn(WO$_4$)$_2$. The third family includes several iron selenides: BaFe$_2$Se$_3$, KFe$_2$Se$_2$, and CaOFeS, which display ferroelectricity or magnetodielectric effect due to exchange striction. The last example is LiFe$_2$F$_6$, which displays electric ferroelectricity due to charge ordering. Its antiferromagnetism can be tuned to ferrimagnetism under moderate compressive strain. Furthermore, it is a candidate material to realize the electric control of magnetization.

As an emerging branch of multiferroics, the iron-based improper ferroelectrics still need extensive researches before the applications. There are some drawbacks regarding their multiferroic performances. For example, the magnetic ordering temperatures are mostly below room temperature even though they have been improved comparing the corresponding iso-structural Mn-based compounds. Furthermore, the magnetic orders in these compounds are typically AFM with nearly zero residue magnetization. To overcome these drawbacks, researchers need to find more candidate compounds with stronger exchange interactions and/or simpler exchange routes, and those iron-based compounds with ferrimagnetism.

\section*{Acknowledgment}
We thank collaborators Q. Z. Huang, V. O. Garlea, T. Zou, S. D. Shen, J. Zhao, E. Dagotto for their contributions in the works reviewed in this article. Work was supported by National Natural Science Foundation of China (Grant Nos. 11674055, 51721001, 11504048, and 11704109), National Key Research and Development Program of China (Grant No 2016YFA0300101), and the Fundamental Research Funds for the Central Universities, China.

\section*{References}
\bibliographystyle{iopart-num}
\bibliography{ref3}
\end{document}